\documentclass{iaus}
\usepackage{graphicx}

\title[The FR Dichotomy] 
{ Evolution of AGN Space Densities and the FR Dichotomy}

\author[Melanie A. Gendre, Jasper V. Wall \& Philip Best]   
{Melanie A. Gendre$^1$, Jasper V. Wall$^1$ \and Philip Best$^2$}

\affiliation{$^1$Department of Physics and Astronomy, University of British
  Colombia, 6224 Agricultural Rd, Vancouver, BC, V6T 1Z1,
  Canada\\
  email: {\tt mgendre@phas.ubc.ca}\\[\affilskip]
  $^{2}$Institute for Astronomy, Royal Observatory, Blackford Hill,
  Edinburgh EH9 3HJ, UK}

\pubyear{2009}
\volume{267}  
\jname{Co-Evolution of Central Black Holes and Galaxies}
\editors{B.M.\ Peterson, R.S.\ Somerville, \& T.\ Storchi-Bergmann, eds.}
\begin{document}

\maketitle

\begin{abstract}
We focus on a comparison of the space densities of FRI and
FRII extended radio sources at different epochs, and find that FRI and
FRII sources show similar space density enhancements in various
redshift ranges, possibly implying a common evolution.
\end{abstract}

\begin{figure}[h]
  \begin{minipage}{4.3cm}
\medskip
\medskip
    \centerline{
      \includegraphics[angle=270,scale=0.18]{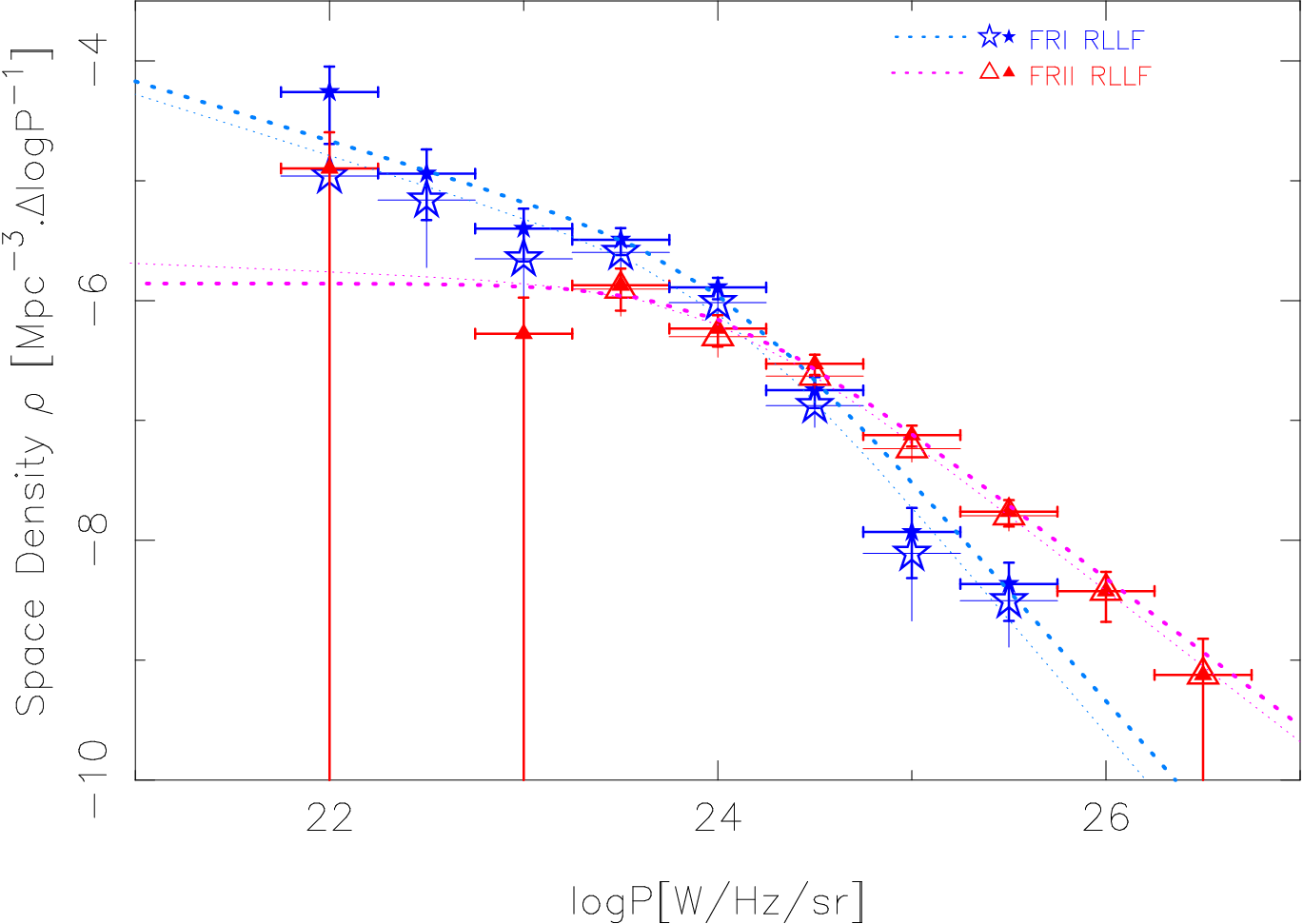}}
\medskip
    \caption[]{\label{RLFI-II}Local luminosity function $\rho(P)$ for
      FRIs and FRIIs, represented by blue stars and red triangles
      respectively.}
  \end{minipage}
  \hfill
  \hspace{0.5mm}
  \begin{minipage}{9cm}
    \begin{minipage}{4.5cm}
      \centerline{
        \includegraphics[angle=270,scale=0.18]{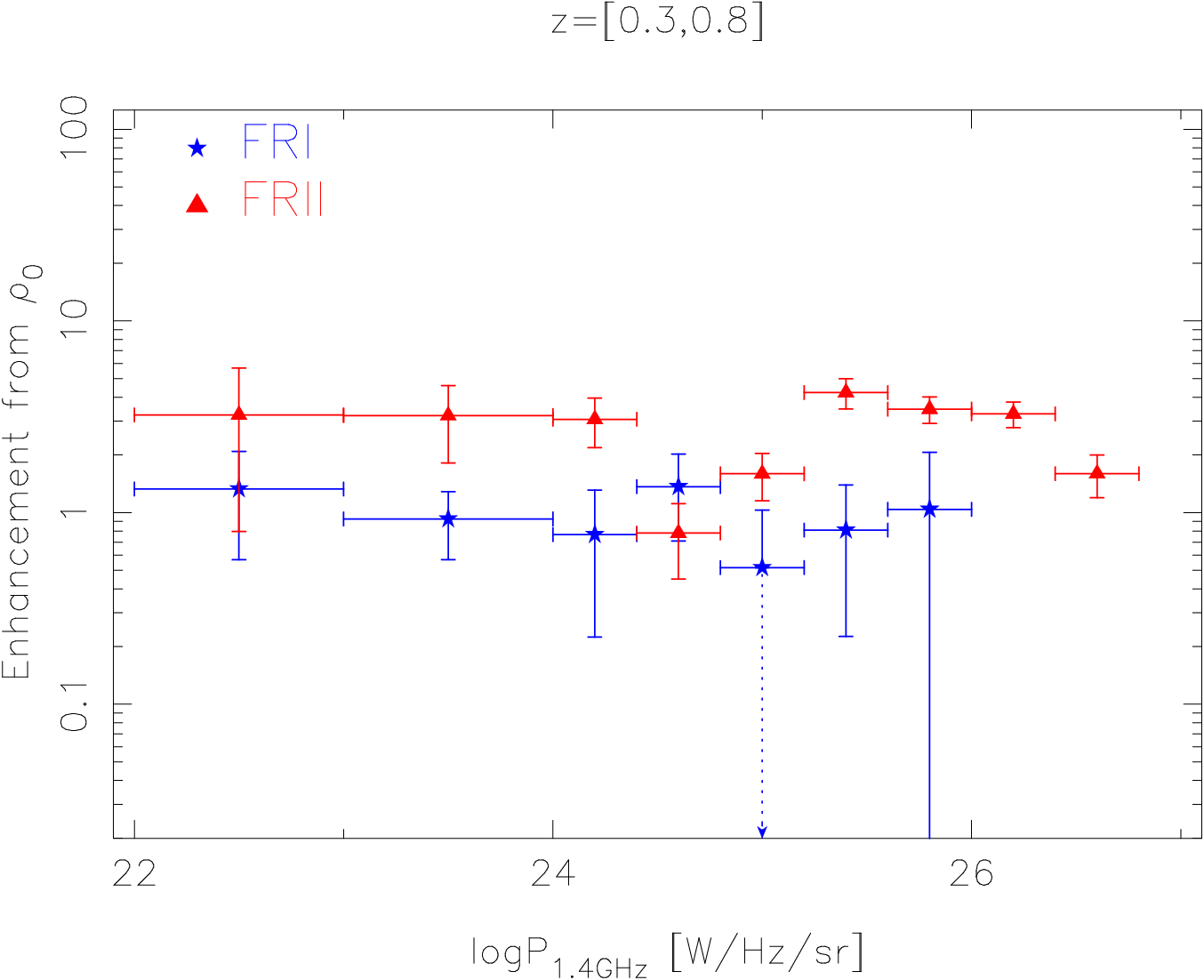}}
    \end{minipage}
    \hfill
    \begin{minipage}{4.5cm}
      \centerline{
        \includegraphics[angle=270,scale=0.18]{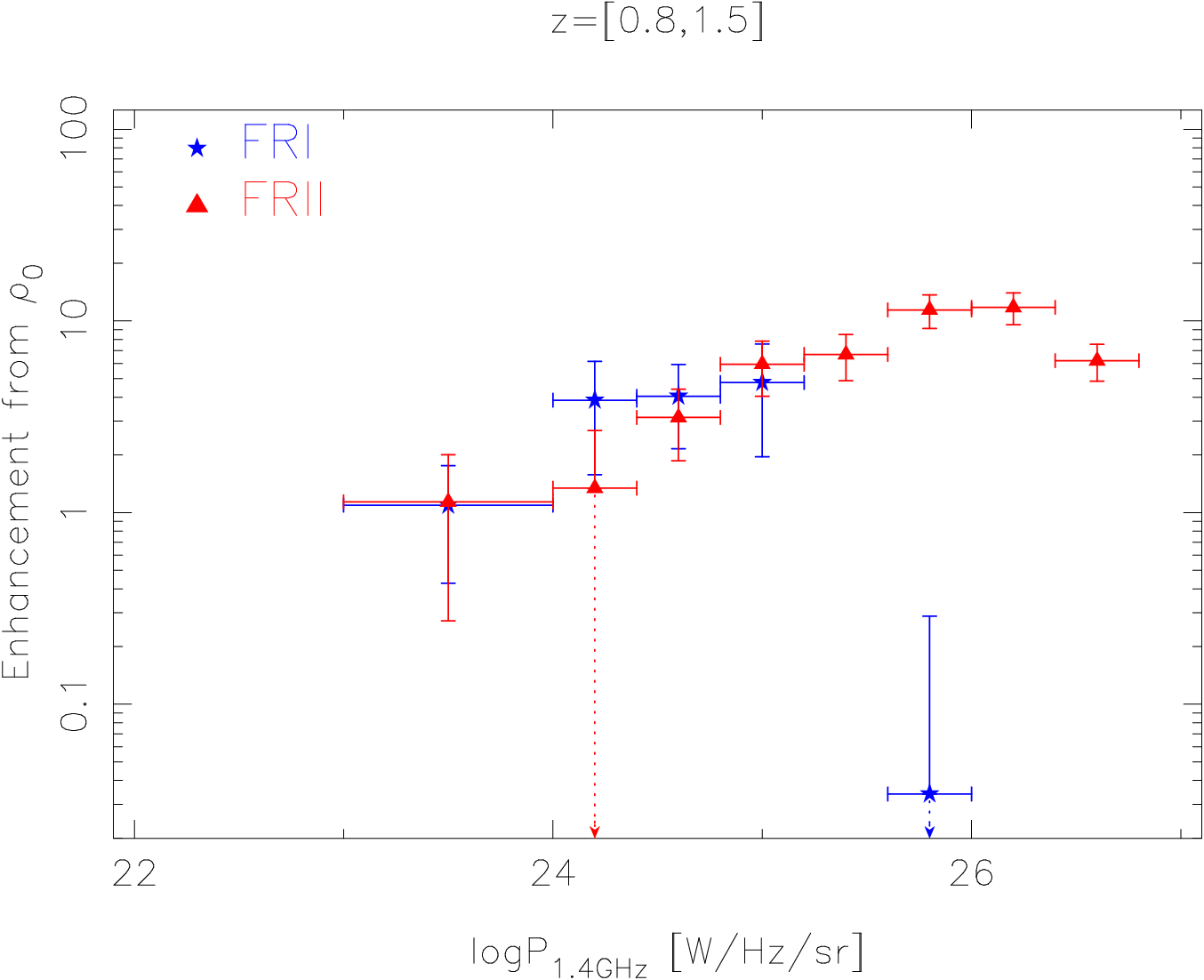}}
    \end{minipage}
    \caption[]{\label{bootFRIvsII}Comparison of the space density
      enhancement between FRI (blue stars) and FRII
      (red triangles) sources, for different redshift bins
      (z=[0.3:0.8] and z=[0.8:1.5]).}
  \end{minipage}

\end{figure}

Based on data compiled in the CoNFIG catalogue \cite[(Gendre \& Wall
  2008; Gendre, Best \& Wall 2009)]{PaperI,PaperII}, we compute the
radio luminosity functions (RLF) for different redshift bins for each
FR (\cite{FR}) population using the $1/V_{max}$ technique. The FRI and
FRII local ($z\le 0.3$) RLFs, in Fig.\ref{RLFI-II}, show apparent
differences. The FRII LRLF does not show any turn-over, suggesting
that there is no sharp luminosity break between FRI and FRII
sources. Overall, these LRLFs indicate that locally FRI and FRII
sources constitute two distinct populations. The RLF for each
population in Fig.~\ref{bootFRIvsII} was then computed in different
redshift bins. The overall behaviour of the enhancement with
luminosity of FRI and FRII sources is very similar. With both
populations show similar enhancement history, there may be a common
mechanism governing the cosmic evolution.

\end{document}